\documentclass[apjl]{emulateapj}

\usepackage{amsmath}
\usepackage{apjfonts}
\usepackage{amssymb}
\usepackage{mathrsfs}
\usepackage{natbib}

\def\newacronym#1#2#3{\gdef#1{#3 (#2)\gdef#1{#2}}}

\newacronym{\cmb}{CMB}{cosmic microwave background}
\newacronym{\isw}{iSW}{integrated Sachs-Wolfe}
\newacronym{\frw}{FRW}{Friedman-Robertson-Walker}  
\def\gw#1{gravitational wave#1 (GW#1)\gdef\gw{GW}}
\def\bh#1{black hole#1 (BH#1)\gdef\bh{BH}}
\def\snr#1{signal-to-noise ratio#1 (SNR#1)\gdef\snr{SNR}}
\def\intisw{I_{\hbox{\tiny{ISW}}}}

\begin{document}


\shorttitle{Gravitational Integrated Sachs-Wolfe Effect}
\shortauthors{P. Laguna et al.}

\title{Integrated Sachs-Wolfe Effect for Gravitational Radiation}

\author{Pablo Laguna\altaffilmark{1}, Shane Larson\altaffilmark{2},
  David Spergel\altaffilmark{3,4}, Nicol\'as Yunes\altaffilmark{5}} 

\altaffiltext{1}{Center for Relativistic Astrophysics and School of
  Physics, School of Physics, Georgia Institute of Technology,
  Atlanta, GA 30332, USA} 
  \altaffiltext{2}{Department of Physics, Utah
  State University, Logan, UT 84322, USA} 
  \altaffiltext{3}{Department
  of Astrophysical Sciences, Princeton University, Princeton, NJ
  08544, USA} 
  \altaffiltext{4}{Princeton Center for Theoretical
  Sciences, Princeton University, Princeton, NJ 08544, USA}
\altaffiltext{5}{Department of Physics, Princeton University,
  Princeton, NJ 08544, USA}

\email{plaguna@gatech.edu}

\begin{abstract}
  Gravitational waves are messengers carrying valuable information
  about their sources. For sources at cosmological distances, the
  waves will contain also the imprint left by the intervening
  matter. The situation is in close analogy with cosmic microwave
  photons, for which the large-scale structures the photons traverse
  contribute to the observed temperature anisotropies, in a process
  known as the integrated Sachs-Wolfe effect. We derive the
  gravitational wave counterpart of this effect for waves propagating
  on a Friedman-Robertson-Walker background with scalar
  perturbations. We find that the phase, frequency and amplitude of
  the gravitational waves experience Sachs-Wolfe type integrated
  effects, this in addition to the magnification effects on the
  amplitude from gravitational lensing. We show that for supermassive
  black hole binaries, the integrated effects could account for
  measurable changes on the frequency, chirp mass and luminosity
  distance of the binary, thus unveiling the presence of
  inhomogeneities, and potentially dark energy, in the Universe.
\end{abstract}

\maketitle


Observations of \gw{s} have a tremendous potential for transforming
our understanding of the cosmos. For sources at cosmological
distances, such as the inspiral of supermassive \bh{} binaries in the
sensitivity window of the proposed Laser Interferometer Space Antenna
(LISA), \gw{s} will not only carry information of the characteristics
of the source but also contain information on the cosmological
expansion of space-time through which these waves propagate, yielding
redshifted measurements of the properties of the binary (e.g.
distance, chirp-mass,
frequency)~\citep{Hughes:2003ty,Finn:1995ah,PhysRevD.57.4535}. When
combined with a coincident electromagnetic (EM) counterpart, \gw{s}
observations from supermassive \bh{} inspirals might have the
potential to serve as standard \emph{sirens} for determining the
distance-redshift relation~\citep{2005ApJ...629...15H}.

In addition to the cosmological expansion effect, \gw{} propagation is
also affected by (1) the proper motion of the source and the receiver
relative to the cosmological flow, (2) the gravitational potentials at
the emitting and receiving locations, and (3) the intervening matter
the \gw{s} traverse. These processes are identical to those
experienced by \cmb{} photons, which lead to temperature anisotropies
$\Delta T/T$ given by
\begin{equation}
\frac{\Delta T}{T}  = \vec n \cdot\vec v|^r_e + \Phi|^r_e + 2\,\int^r_e \partial_{\tau} \Phi
d\lambda\,,
\label{eq:sw}
\end{equation} 
where $\Phi$ is a scalar perturbation, $v$ is the source velocity,
$\vec{n} = \vec{x}/r$ is a unit vector, and the sub- and super-scripts
$r$ and $e$ stand for evaluation at the receiver and emission
location, respectively. In Eq.~(\ref{eq:sw}), the first term is the
Doppler correction; the second term gives the ordinary Sachs-Wolfe
effect~\citep{Sachs:1967er}; and the third is the \isw{} or Rees-Sciama
effect~\citep{Rees:1968zz}. The \isw{} term is an integral along the
photon geodesic, with affine parameter $\lambda$, which accounts for a
net gain or loss in the temperature of the \cmb{} photons due to
changes in the gravitational potential of scalar inhomogeneities. This
effect has proven to be a crucial in interpreting observations of
\cmb{} inhomogeneities (see
eg.~\citep{Giovannini:2004rj,Afshordi:2004kz} for a review).

In this \emph{Letter}, we present a derivation of the \gw{}
counterpart to the EM \isw{} effect. We focus on \gw{s} propagating on
a \frw{} background with generic scalar perturbations. The derivation
follows Isaacson geometric optics
approximation~\citep{Isaacson:1968ra,Isaacson:1968gw}. We find that the
\gw{} phase and frequency experience changes similar to those in
Eq.~(\ref{eq:sw}) for \cmb{} photons as the former propagates on a
perturbed FRW background. In addition, we derive the corresponding
changes to the \gw{} amplitude, including magnification effects due to
gravitational lensing. We estimate the impact of these integrated
effects on measurements of the chirp mass, frequency and luminosity
distance of supermassive \bh{s} at cosmological distances and discuss
their potential for unveiling the presence of inhomogeneities in the
Universe. Latin letters from the beginning and middle of the alphabet
will denote spacetime and spatial indices, respectively. We use
geometric units $G = c = 1$.


\emph{Isaacson's Geometric Optics Approximation.} Following
Isaacson~\citep{Isaacson:1968ra,Isaacson:1968gw}, the
space-time metric is decomposed as $ g_{a b} = \gamma_{ab} + \epsilon h_{ab}, $
where $\gamma_{ab}$ is a background metric and $h_{ab}$ a \gw{} metric
perturbation.  $\epsilon$ is a book-keeping parameter to keep track of the order
of the perturbation; formally $\epsilon \sim \lambda/L \ll 1$,  with $\lambda$ the wavelength of radiation and $L$ the radius of curvature of the
background. At the end 
of the calculation, $\epsilon$ is set to unity.  
$\gamma_{ab}$ is in addition decomposed as 
$\gamma_{ab} = \gamma^{(0)}_{ab} + \gamma^{(1)}_{ab} $, 
where the labels in parenthesis denote the order of the perturbation. 
In our case, 
$\gamma^{(0)}_{ab} = a^2(\tau)\eta_{ab}$ is the flat \frw{} metric and
$\gamma^{(1)}_{ab} = a^2(\tau)\delta\eta_{ab}$ its first order perturbation. 
We consider only scalar perturbations $(\Phi,\Psi)$, such that $
\delta\eta_{ab}dx^adx^b = -2\,\left[ \Phi\,d\tau^2 +
  \Psi\,\eta_{ij} dx^i dx^j \right]$.

The linearized Einstein equations to first order in $\epsilon$ and in Lorentz-gauge
($\nabla_b \bar h^{ab} = 0$) are given by:
\begin{eqnarray}
  &&  \nabla_c\nabla^c \bar h_{ab} + 2\,R_{cadb}\bar h^{cd} =
  -16\,\pi\,{\cal T}_{ab}\nonumber \\
  && - 8\,\pi\left[
    \bar h_{cd}T^{cd}\gamma_{ab} + \bar h_{cd} \gamma^{cd}
\left( T_{ab} -\frac{1}{2} T\gamma_{ab}\right)\right]\,,
\label{eq:hbox}
\end{eqnarray}
In Eq.~(\ref{eq:hbox}), $R_{abcd}$, $T_{ab}$ and $\nabla_a$ denote respectively the
curvature tensor, the stress-energy tensor and the covariant
derivative associated with the background metric $\gamma_{ab}$. ${\cal
  T}_{ab}$ is the stress-energy tensor associated with $h_{ab}$, whose
trace-reversed form is $\bar h_{ab} = h_{ab} - \gamma_{ab} h/2$, where
$h$ and $T$ are traces with respect to the background. The
stress-energy tensor $T_{ab}$ is decomposed as $T_{ab} =
T^{(0)}_{ab} + T^{(1)}_{ab}$ with $T^{(0)ab} =
(\rho+p)U^{(0)a}U^{(0)b}+ p\gamma^{(0)ab}$ and $T^{(1)00} =
a^{-2}(\delta\rho- 2\,\rho\,\Phi)$, $T^{(1)0j} =
a^{-2}(p+\rho)v^j$, and $T^{(1)ij} = a^{-2}(\delta
p+2\,p\,\Psi)\,\eta^{ij}$. Similarly, the four-velocity of the fluid
in the background geometry is given by $U^a = U^{(0)a} + U^{(1)a}$
with $U^{(0)a} = (1,\vec 0)/a$ and $U^{(1)a} = (-\Phi,v^i)/a$. The
quantities $\rho$ and $p$ are the density and pressure of the
background fluid, while $\delta \rho$ and $\delta p$ are the density
and pressure perturbations, where $v^{i}$ is the $3$-velocity of the
perturbed fluid.

The field equations can be simplified by choosing the TT gauge in this
local Lorenz frame: $\bar h_{00} = \bar h_{0j} = 0$ and $\bar
h_{ij}\eta^{ij} = 0$. Thus, $\bar h_{ab}\gamma^{ab} = 0$ and
$T^{ab}\bar h_{ab} = 0$. In addition, we neglect the response of the
matter background to the presence of the \gw{} $\bar h_{ab}$ and set
${\cal T}_{ab} = 0$. Therefore, Eq.~(\ref{eq:hbox}) becomes:
\begin{equation} \nabla_c\nabla^c \bar h_{ab} + 2\,R_{cadb}\bar h^{cd}
  = 0. \label{eq:hbox2} \end{equation}

Under Isaacson's~\citep{Isaacson:1968ra,Isaacson:1968gw} shortwave or
geometric optics approximation, the \gw{} can be written as
\begin{equation}
\label{ansatz}
\bar{h}_{ab} = A_{ab} \; e^{i \phi/\epsilon} = e_{ab} \;{\cal A}\;e^{i \phi/\epsilon} = e_{ab} h\,,
\end{equation}
where ${\cal A}$ and $\phi$ are real functions of retarded time $u =
\tau - r$, $e_{ab}$ is a polarization tensor and $r$ is the distance to the
source. 

Isaacson's approximation allows us to simplify the field equations dramatically. 
With Eq.~(\ref{ansatz}), Eq.~(\ref{eq:hbox2}) becomes
\begin{eqnarray}
&&\epsilon^{-2}[-k^ck_cA_{ab}] + i\epsilon^{-1}[2\,k^c\nabla_c A_{ab} +
A_{ab}\,\nabla_ck^c]\nonumber\\
&&+ [\nabla^c\nabla_c A_{ab} + 2\,R_{cadb}A^{cd}] = 0\,,
\label{eq:optics}
\end{eqnarray}
where $k_a = \nabla_a\phi$ is the \gw{} wave-vector. 
To ${\cal{O}}(\epsilon^{-2})$, Eq.~\eqref{eq:optics} requires that $k^ak_a =
0$, implying that \gw{} rays are null vectors and
the curves $x^a(l)$, defined by $dx^a/dl = k^a$, are null
geodesics, i.e.~$k^b\nabla_bk^a = 0$. To ${\cal{O}}(\epsilon^{-1})$, 
Eq.~(\ref{eq:optics}) implies that $k^c\nabla_c
e_{ab} = 0$, i.e.~the polarization tensor is parallel-transported
along null geodesics, and thus
\begin{equation}
\label{WKB:2}
\frac{d}{dl}\ln {\cal{A}}  = - \frac{1}{2} \nabla_ak^a \,,
\end{equation}
where we have used $d/dl \equiv k^a\nabla_a$. Eq.~(\ref{WKB:2}) shows that
the \gw{} amplitude decreases as the null rays diverge. This equations
can also be rewritten as $\nabla_a( {\cal{A}}^2\,k^a) = 0$.  


\emph{Integrated Sachs-Wolfe or Rees-Sciama effect.} Since the
\gw~wave-vector $k^a$ satisfies the null geodesic equation, one can
essentially follow the derivation of the \isw{} effect for
\cmb{} photons as given by \citet{1996PhRvD..53.2920P}.  The first
step is to notice that the null geodesics $x^a(l)$ with affine
parameter $l$ of \gw{s} in the background metric $\gamma_{ab} =
a^2(\eta_{ab}+\delta\eta_{ab})$ are the same as the null geodesics
$\tilde{x}^a(\lambda)$ with affine parameter $\lambda$ in the perturbed
Minkowski metric $\tilde\gamma_{ab} = \eta_{ab}+\delta\eta_{ab}$. The
affine parameters, metrics and wave-vectors are related by $dl =
ad\lambda$, $\gamma_{ab} = a^2\tilde\gamma_{ab}$ and $k^a =
a^{-2}\tilde k^a$, respectively.

We will set coordinates such that the observer is at the end of the
\gw{} world-line, receiver's location $x^a(\tau_{r}) = (\tau_r ,\vec 0)$, with
the world-line starting at the ``surface'' of emission defined by the
spacelike hypersurface of constant conformal time $\tau_e$. The
perturbed null geodesic $\tilde{x}^a(\lambda) = \tilde{x}^{(0)a}(\lambda) +
\tilde{x}^{(1)a}(\lambda)$ and its corresponding wave-vector $\tilde
k^a(\lambda) = \tilde k^{(0)a}(\lambda) + \tilde k^{(1)a}(\lambda)$.
The lowest order contributions are given by $\tilde{x}^{(0)a}(\lambda) =
[\lambda,(\lambda_r-\lambda) \,n^i]$ and $\tilde k^{(0)a} = (1,
-n^i)$ , where the vector $n^i$ points in the sky direction of arrival of
the \gw{.}

The next order wave-vector can be obtained from the null geodesic
equation associated with the perturbed Minkowski metric
$\tilde{\gamma}_{ab}$: 
\begin{equation}
  \frac{d}{d\lambda}\tilde k^{(1)a} +
  \widetilde\Gamma^{(1)a}_{bc}\tilde k^{(0)b}\tilde k^{(0)c}= 0\,,
  \label{eq:geo} 
\end{equation} 
where we have used that $\widetilde\Gamma^{(0)a}_{bc} = 0$. The time
component of Eq.~(\ref{eq:geo}) yields
\begin{equation} \frac{d}{d\lambda}\tilde k^{(1)0}
  = \partial_{\tau}(\Psi + \Phi) - 2\,\frac{d\Phi}{d\lambda} \,,
  \label{eq:k1dot} 
\end{equation}
where $d\Phi/d\lambda \equiv \partial_{\tau}\Phi + \tilde k^{(0)i}\partial_i\Phi$ 
and its integration yields
\begin{equation} \tilde k^{(1)0}=
  -\left.(\Phi+\Psi)\right|_{\lambda_e}-
  2\,\Phi |^{\lambda}_{\lambda_e} + \intisw\, \label{eq:k10}
\end{equation}
where we have defined
\begin{equation} \intisw \equiv
  \int^{\lambda}_{\lambda_e}\partial_{\tau}(\Psi + \Phi)d\lambda' \,.
\end{equation}
The spatial component of Eq.~(\ref{eq:geo}) yields
\begin{eqnarray}
\frac{d}{d\lambda} \tilde k^{(1)i}_\parallel 
&=& \tilde k^{(0)i}\left[\frac{d }{d\lambda} (\Psi-\Phi) + \partial_{\tau}(\Phi+\Psi)\right]\,,
\label{eq:k1paradot}
\\
\frac{d}{d\lambda} \tilde k^{(1)i}_\perp &=& -(\eta^{ij}-\tilde
k^{(0)i}\tilde k^{(0)j})\partial_j(\Phi+\Psi)\,,
\label{eq:k1perpdot}
\end{eqnarray}
where we have introduced the notation $\tilde k^{(1)i}_{\parallel} =
\tilde k^{(0)i}\tilde k_j^{(0)} \tilde k^{(1)j}$ and $\tilde
k^{(1)i}_\perp = \perp^i_j \tilde k^{(1)j}$, such that $\tilde k^{(1)i}
= \tilde k^{(1)i}_\parallel + \tilde k^{(1)i}_\perp$.
The perpendicular operator $\perp^i_j =
\delta^i_j-\tilde k^{(0)i}\tilde k_j^{(0)}$ projects components of tensors 
orthogonal to the unperturbed wave-vector $k^{(0)i}$, and thus, 
``parallel'' and ``perpendicular'' are operations defined with respect to this 
vector.
Integration of Eqs.~(\ref{eq:k1paradot}) and (\ref{eq:k1perpdot}) yields
\begin{eqnarray}
  \tilde k^{(1)i}_\parallel &=& \tilde k^{(0)i}  \left[
    \left.(\Psi-\Phi)\right|^{\lambda}_{\lambda_e}  + \intisw\right]\,,\label{eq:k1ipara}\\
  \tilde k^{(1)i}_\perp&=& -\perp^{ij}\int^{\lambda}_{\lambda_e}\partial_j(\Phi+\Psi)d\lambda' \,,\label{eq:k1iperp}
\end{eqnarray}
where the integration constants in Eqs.~(\ref{eq:k10}), (\ref{eq:k1ipara}) and
(\ref{eq:k1iperp}) are chosen so that $\tilde k^a$ is null at $\lambda_{e}$.

The \gw{} phase $\phi$ is obtained from 
\begin{equation} 
\frac{d}{d\lambda}\phi = \tilde k^{(0)a}\nabla_a\phi 
  = -\tilde k^{(1)0} + \tilde k_i^{(0)} \tilde k^{(1)i}_\parallel = \Psi+\Phi \,,
\end{equation} 
which after integration yields 
\begin{equation}
\label{eq:phase}
\delta\phi = \phi - \phi_e = \int^{\lambda}_{\lambda_e}(\Psi+\Phi)d\lambda'\,.
\end{equation}
This phase shift corresponds to the Shapiro time delay commonly
associated with photon propagation.  

The \gw{} frequency $\omega$ in the reference frame of the
cosmological fluid defined by the four-velocity $U^a = (1-\Phi, v^i)/a
$ is given by $\omega = - U^a k_a = a^{-1}
[1-\Psi({\lambda_e})-\Phi|^{\lambda}_{\lambda_e} + \vec n\cdot\vec v +
\intisw]\,$. From this, one finds that the emitting and receiving
frequencies are related via 
\begin{equation}
\frac{\omega_r}{\omega_e} = \frac{f_r}{f_e} =
\frac{(1-\chi)}{(1+z)}
\label{eq:freqratio}
\end{equation}
where $\chi = \Phi |^r_e-\vec n \cdot\vec v|^r_e - \intisw(\lambda_r)$
and $a_r/a_e = 1 + z$. Notice that the redshifted frequency acquires an \isw{}
correction, as well as a non-integrated Doppler one. 

The \gw{} amplitude ${\cal{A}}$ is obtained from Eq.~(\ref{WKB:2}).
To lowest order in the scalar perturbations, we have
\begin{equation}
  \frac{d}{d\lambda}\ln ({\cal A}^{(0)}\,a) = -\frac{1}{2}\widetilde\nabla^{(0)}_a \tilde k^{(0)a}\,.
\label{eq:Q0}
\end{equation}
which implies that ${\cal A}^{(0)}\,a\,r = {\cal Q}$ is constant along
the null geodesic with $r$ the areal coordinate distance defined by
the background metric $\eta_{ab}$. The quantity ${\cal Q}$ is
determined by the local wave-zone source solution. At the receiving
location, ${\cal Q}$ is given by the same solution evaluated at the
retarded time.

To next order in the scalar perturbations, Eq.~(\ref{WKB:2}) yields
\begin{equation}
-2\,\frac{d}{d\lambda}\xi = \partial_{\tau}\tilde k^{(1)0} +\partial_i\tilde k^{(1)i}_{\parallel}  + \partial_i\tilde k^{(1)i}_{\perp} 
+ \widetilde\Gamma^{a(1)}_{ab}\tilde k^{(0)b}\,,
\label{eq:dxi}
\end{equation}  
where we have introduced ${\cal A} = {\cal A}^{(0)}(1+\xi)$ and
each term is given by
\begin{eqnarray}
\partial_{\tau}\tilde k^{(1)0}  &=& \partial_{\tau}(-2\,\Phi + \intisw) \nonumber \\
\partial_i\tilde k^{(1)i}_{\parallel}
&=& \frac{d}{d\lambda}(\Psi-\Phi+\intisw) -
\partial_{\tau} (\Psi-\Phi+\intisw)\nonumber \\
\partial_i\tilde k^{(1)i}_{\perp}  &=&-\perp^{ij}\int^{\lambda}_{\lambda_e}\partial_{ij}(\Phi+\Psi)d\lambda' \nonumber \\
\widetilde\Gamma^{(1)a}_{ab}\tilde k^{(0)b} &=& \frac{d}{d\lambda}
(\Phi-3\,\Psi)\,. \nonumber
\end{eqnarray}
Above, we have ignored terms of ${\cal O}(r^{-1})$ since the $1/r$
dominant dependence in ${\cal A}$ has been explicitly accounted at the
lowest order. Equation~(\ref{eq:dxi}) can be written as
\begin{eqnarray}
2\,\frac{d}{d\lambda}\xi &=& -\partial_{\tau}(\Psi+\Phi)+ \frac{d}{d\lambda}(-2\,\Psi+\intisw)\nonumber\\
&-&\perp^{ij}\int^{\lambda}_{\lambda_e}\partial_{ij}(\Phi+\Psi)d\lambda' \,,
\end{eqnarray}
which after integration yields 
\begin{equation}
\xi =  -\Psi|^{\lambda}_{\lambda_e} -\frac{1}{2}\perp^{ij}\int^{\lambda}_{\lambda_e}\int^{\lambda'}_{\lambda_e}\partial_{ij}(\Phi+\Psi)d\lambda'' d\lambda'\,.
\label{eq:xi}
\end{equation}
Notice that the \isw{} contribution to the amplitude has canceled. The
remaining integral contribution is the magnification due to
gravitational lensing~\citep{2006ApJ...644...80T}.

Combining all results, the \gw{} takes the form
\begin{equation}
h = {\cal A}\;e^{i\phi} = \frac{{\cal Q}(1+z)}{d_L}(1+\xi)e^{i(\phi_e+\delta\phi)}\,,
\label{eq:hab}
\end{equation}
where we have set $\epsilon$ to unity, $\delta\phi$ and $\xi$ are given by Eqs.~(\ref{eq:phase}) and
(\ref{eq:xi}), respectively, and $d_L = ar(1+z)$ is the luminosity
distance. At Newtonian (quadrupole) order and for an inspiraling
binary~\citep{1995PhRvD..52..848P}, ${\cal
  Q} = {\cal M}_e(\pi\,f_e\, {\cal M}_e)^{2/3}$ and $\phi_e  = \phi_c
-(\pi\,f_e\, {\cal M}_e)^{-5/3}/16$, with ${\cal M}_e$ and
$f_e$ the intrinsic ``chirp mass'' and frequency of the binary,
$\phi_c$ the value of the phase at $f=\infty$ and 
$t(f) = t_c - (5/256) {\cal M}_e (\pi\,f_e\, {\cal M}_e)^{-8/3}$.
Therefore, Eq.~(\ref{eq:hab}) becomes
\begin{equation}
h = \frac{{\cal
    M}_e(1+z)}{d_L}(\pi\,f_e\, {\cal
  M}_e)^{2/3}(1+\xi)e^{i(\phi_e+\delta\phi)}\,.
\label{eq:he}
\end{equation}
The modified redshift relation [Eq.~(\ref{eq:freqratio})] implies that $f_e
{\cal M}_e = f_r {\cal M}_e (1+z)/(1-\chi)$, and thus ${\cal M}_r = {\cal
  M}_e (1+z)/(1-\chi)$ and Eq.~(\ref{eq:he}) becomes
\begin{equation}
h = \frac{{\cal M}_r}{D_L}(\pi\,f_r\, {\cal M}_r)^{2/3}e^{i\phi_r}\,.
\label{eq:he2}
\end{equation}
where $D_L = d_L/ (1-\chi+\xi)$ is the modified luminosity distance
and $\phi_{r}$ is the modified version of Eq.~\eqref{eq:phase}.

The Fourier transform of Eq.~(\ref{eq:he2}), using the stationary
phase approximation, is $\tilde h = ({\cal M}^2_r/D_L) (f_r\, {\cal
  M}_r)^{-7/6} e^{i\psi_r}\,,$ where we have neglected the
antenna-pattern functions and where $\psi_r \equiv 2\,\pi\,f_{r}\,t_o
+ \phi_r(t_o)$ with $t_o$ a stationary point of the phase.  The square
of the \snr{}, $\sigma^2 = 4\int^\infty_0 |\tilde h|^2/S_n df_{r}$, is
then given by
\begin{equation}
\sigma^2
= 4\left(\frac{{\cal M}_r}{D_L}\right)^2\int^\infty_0  \frac{(f_r\, {\cal M}_r)^{-7/3}}{(S_n/{\cal M}_r)}\,d(f_r\,\, {\cal M}_r),
\end{equation}
with $S_n$ the spectral noise density.

Therefore, the changes to the chirp mass, frequency, luminosity
distance and \snr{} induced by the scalar inhomogeneities in the
background are given by
\begin{equation}
\frac{\delta {\cal M}}{ {\cal M}_z} = -\frac{\delta f}{f_z} = \chi\,,\qquad
\frac{\delta D_L}{d_L} =  \chi-\xi\,,\qquad
\frac{\delta \sigma}{\sigma}  =  \frac{\chi}{2}+\xi\,,
\end{equation} 
where ${\cal M}_z = {\cal M}_e (1+z)$ and $f_z = f_e/(1+z)$.

\emph{Root-Mean-Squared Fluctuations.} Next, we expand $\chi$ and
$\xi$ in Fourier modes to relate them with the density perturbation
modes: $\Phi_{k} = \Psi_{k} = - 3 H_{0}^{2} \delta_{k}^{(0)} D(t)/ (2
k^{2} a)$, where $D(t)$ is the linear growth function and
$\delta_{k}^{(0)}$ is the Fourier coefficient of the density
perturbation at zero redshift. The ensemble average is performed over
the density perturbations via the definition of the power spectrum
\begin{equation}
\left< \delta_{k}^{(0)} \delta_{k'}^{\star}{}^{(0)} \right> = \delta^{3}(k - k') \left(2 \pi\right)^{3} P_{\delta}^{(0)}(k) = \frac{4}{25} \frac{k^{4}}{H_{0}^{4}} P_{R} T^{2} 
\end{equation}
where $P_{R}(k) = 2 \pi^{2} \Delta_{R}^{(0)2} k^{-3}$ is the power
spectrum of curvature perturbations, with $\Delta_{R}^{(0)} = 2.445
\times 10^{-9}$, and $T(k)$ is a transfer function, which we
approximate via the fitting relations given
in~\citep{Eisenstein:1997jh}.  The $e^{i k \cdot x}$ factor that arises
in the Fourier transform is expanded in spherical Bessel 
functions and the integration over these functions is
performed through the Limber approximation~\citep{LoVerde:2008re}
\begin{equation}
\int dt' \frac{f(t')}{\sqrt{t'}} J_{\ell + 1/2}(k t') \sim \frac{1}{\sqrt{k (\ell + 1/2)}} f\left(\frac{\ell+1/2}{k}\right),
\end{equation}
which requires that $\ell + 1/2 = k t'$, valid for large $\ell$.

With these considerations in mind,
\begin{eqnarray}
\left<\chi^{2}\right> &\sim& \sqrt{\frac{18}{25}} \Omega_{m,0} \left[\int_{k_{min}}^{\infty} k dk P_{R}\right]^{1/2} \left[ \int_{0}^{t} dt' \dot{g}^{2}(t') \right]^{1/2}
\\
\left<\xi^{2}\right> &\sim& \sqrt{\frac{18}{25}} \Omega_{m,0} \left[\int_{k_{min}}^{\infty} k^{5} dk P_{R}\right]^{1/2} \left[ \int_{0}^{t} dt' t'^{2} g^{2}(t') \right]^{1/2}
\end{eqnarray}
where $k_{min} \approx 0.0004$, $g(t) \equiv D(t)/a(t)$
and $\Omega_{m,0} \approx 0.27$. The $
\dot{g}$ is approximated via the fitting function $\dot{D}/D =
\Omega_{m,0}^{\gamma} H_{0}^{2\gamma} a^{-3 \gamma} H^{-2\gamma + 1}$~\citep{Acquaviva:2008qp,2008PhLB..660..439P},
where $\gamma \approx 0.557 - 0.002 z$, $H$ is the Hubble parameter, and $H_{0}$ 
is its value today. $a(t)$ and $H(t)$ are obtained by solving the
Friedman equations numerically for a cosmology with parameters
measured in~\citep{Komatsu:2008hk}.

\begin{figure}
 \begin{center}
 \begin{tabular}{cc}
      \includegraphics[width=70mm]{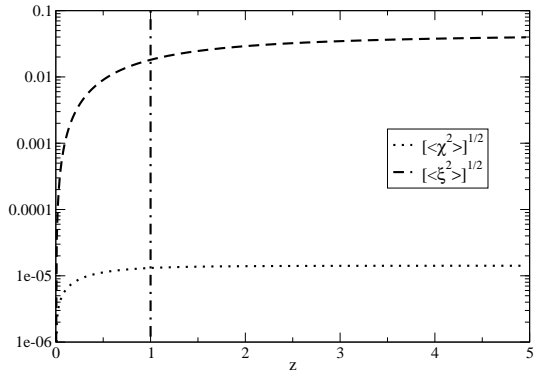} 
 \end{tabular}
 \caption{\label{figure} Root-mean-squared fluctuation of $\chi$
   (dotted) and $\xi$ (dashed) as a function of redshift.}
\end{center}
\end{figure}

Figure~\ref{figure} shows the root-mean-squared fluctuation of $\chi$
(dotted) and $\xi$ (dashed) as a function of redshift. The vertical
line approximately corresponds to the limit to which the Advanced
Laser Interferometer Gravitational Observatory (LIGO) will be able to
see for an optimally-oriented binary. We have here 
somewhat underestimated the $\left<\chi^{2}\right>^{1/2}$ term, as have neglected
non-linear corrections to the power spectrum due to galaxy and cluster 
formation~\citep{Peacock:1996ci}.

\emph{Data Analysis Implications.}  The detection of \gw{s} by LISA or
any other instrument will not be affected by matter inhomogeneities,
since the phase correction accounts for a total phase shift that is
extremized over during \gw{} extraction.  Parameter estimation,
however, will clearly be affected by matter inhomogeneities.  LISA is
expected to be sensitive to the chirp mass and the luminosity distance
to $\Delta \ln {\cal{M}} \sim 10^{-6}$ and $\Delta \ln D_{L} \sim
10^{-3}$~\citep{Arun:2007hu} for low-redshift sources ($ z \sim 0.55$), but also 
sensitive to high redshift sources ($z\sim5-10$) up to $\Delta \ln D_{L} \sim
10^{-2}$~\citep{2002MNRAS.331..805H,2004PhRvD..70d2001V}. Our calculations
confirm that $\xi$ (associated with weak-lensing) will be a noise source 
in the use of standard sirens to measure the equation of state of dark energy 
through the redshifted luminosity distance~\citep{2006ApJ...637...27K}.  

Alternatively, one could view these effects as a new link between EM measurements
of density inhomogeneities and LISA observations. In order to achieve this goal, however,
one would first have to break the degeneracy between $({\cal{M}}_{z}, z, \chi)$
or $(d_{L}, z,\xi, \chi)$. Given a coincident EM and GW detection, one might be able to achieve just this, 
by electromagnetically determining the redshift and the component masses 
via host galaxy identification and correlations between galaxy luminosity and BH mass.  Another
possibility would be to use large scale structure observations to measure $\delta_{k}$
and predict $\chi$. Cosmologists already use large scale structure observations to predict
the \isw{} term for the CMB~\citep{Afshordi:2004kz}.
Such measurements would then open up, for the first time, studies of
cross-correlations between \gw{s} and large-scale structure surveys of
dark matter~\citep{1996PhRvL..76..575C} and possibly dark
energy~\citep{2004PhRvD..69b7301C}. Furthermore, a detection of the
cross-correlation between matter distribution and the GW \isw{} effect
could potentially be another test of GR since it would show that GWs
propagate in the same metric as EM radiation.

\acknowledgements We thank Viviana Acquaviva, Frans Pretorius and
Scott Hughes for useful comments and suggestions. Work supported in
part by NSF grants PHY-0855892, PHY-0903973, PHY-0941417 and PHY-0914553.


\end{document}